\documentclass[a4paper]{article}

\usepackage{INTERSPEECH2020}
\usepackage{url}

\title{Speaker dependent acoustic-to-articulatory inversion \\ using real-time MRI of the vocal tract}
\name{Tam\'as G\'abor Csap\'o$^{1,2}$}
\address{
  $^1$Department of Telecommunications and Media Informatics, \\
	Budapest University of Technology and Economics, Budapest, Hungary \\
	$^2$MTA-ELTE Lend\"ulet Lingual Articulation Research Group, Budapest, Hungary}
\email{csapot@tmit.bme.hu}

\begin{document}

\maketitle
\begin{abstract}
  
Acoustic-to-articulatory inversion (AAI) methods estimate articulatory movements from the acoustic speech signal, which can be useful in several tasks such as speech recognition, synthesis, talking heads and language tutoring. Most earlier inversion studies are based on point-tracking articulatory techniques (e.g. EMA or XRMB). The advantage of rtMRI is that it provides dynamic information about the full midsagittal plane of the upper airway, with a high 'relative' spatial resolution.
In this work, we estimated midsagittal rtMRI images of the vocal tract for speaker dependent AAI, using  MGC-LSP spectral features as input. We applied FC-DNNs, CNNs and recurrent neural networks, and have shown that LSTMs are the most suitable for this task.
As objective evaluation we measured normalized MSE, Structural Similarity Index (SSIM) and its complex wavelet version (CW-SSIM). The results indicate that the combination of FC-DNNs and LSTMs can achieve smooth generated MR images of the vocal tract, which are similar to the original MRI recordings (average CW-SSIM: 0.94). 

\end{abstract}

\section{Introduction}

Articulation is directly linked with the acoustic speech signal in the speech production process.
The acoustic-to-articulatory inversion (AAI) methods estimate articulatory movements from the acoustic speech signal~\cite{Richmond2002}. Recently, there has been a significant interest in AAI, because learning the correlation between articulation and acoustics could improve the performance of several tasks such as speech recognition~\cite{King2007}, synthesis~\cite{Ling2009} and talking heads~\cite{Wang2010}. It can help the visualization of speech production as 3D articulatory animations for pronunciation training and language tutoring~\cite{Katz2014}.

Statistical mapping techniques are suitable for the conversion of articulatory movements into speech and vice versa~\cite{Hueber2013inv,Hueber2012,Hueber2015,Csapo2017c,Moliner2019}. Both speaker dependent (SD-AAI) and independent (SI-AAI) approaches are available~\cite{Illa2018}. Several methods have been proposed to tackle the SD-AAI problem including codebooks~\cite{Ouni2005}, Gaussian Mixture Models (GMM)~\cite{Toda2008,Sepulveda2013}, Hidden Markov Models (HMM)~\cite{Zhang2008}, and Mixture Density Networks~\cite{Richmond2006}. Furthermore, during the past few years, researchers started to use Deep Neural Networks (DNN)~\cite{Uria2012,Wu2015,Liu2015}, convolutional~\cite{Illa2019b} and recurrent networks~\cite{Illa2019}. The lowest error in predicting articulatory position was achieved with the combination of acoustic and textual input, using a bottleneck long-term recurrent convolutional neural network (BLTRCNN)~\cite{Yu2018}.

Definitely, all these approaches need parallel acoustic-articulatory data for training the AAI model. Hence, most above inversion studies are based on Electromagnetic Articulography (EMA) or X-ray Microbeam (XRMB) data, which can track only several points of the articulatory organs, and therefore provide limited input information. Compared to EMA and XRMB, imaging methods (e.g.\ UTI: Ultrasound Tongue Imaging and MRI: Magnetic Resonance Imaging) have the advantage that the tongue surface is fully visible~\cite{Csapo2017c,Moliner2019,Ramanarayanan2018,Toutios2019}. The typical result of 2D ultrasound and 2D MRI recordings is a series of gray-scale mid-sagittal images in which the tongue surface contour has a greater brightness than the surrounding tissue and air.
We started to use raw scanline tongue ultrasound data for SD-AAI using feedforward DNNs, with 25-dimensional MFCC as input~\cite{Porras2019}. The results indicated that already with a simple DNN consisting of two hidden layers (1000 neurons each) relatively good accuracy could be achieved. These initial results showed that SD-AAI is feasible using ultrasound tongue imaging -- but until now, only simple DNNs were used for this task.

Recently, significant advances in MR research (software, hardware, and reconstruction strategies) have allowed real-time MRI (rtMRI) to be a powerful modality for speech production research and for investigating the movement of the articulators~\cite{Ramanarayanan2018,Toutios2019,Narayanan2014}. The advantage of rtMRI is that it provides dynamic information about the full midsagittal plane of the upper airway, even during continuous spoken utterances. It can capture not only lingual, labial, and jaw motion but also the articulatory motion of the velum and the pharyngeal region, which is typically not possible with other articulatory acquisition techniques. Besides, such imaging data helps to comprehend the generation of coronal, pharyngeal, and nasal segments. The sampling rates of rtMRI are relatively low (around 20--25~fps), but are acceptable for running speech. A disadvantage is the large background noise in speech recordings, but noise cancellation can yield an acceptable speech signal, which is synchronized to the articulatory signal. Also, the presence of a substantial number of artifacts and noise make automatic extraction and interpretation of features a difficult problem. Overall, rtMRI provides high relative spatial information in the midsagittal view with relatively low temporal resolution~\cite{Toutios2019}; therefore it is a potentially suitable technique for the target of acoustic-to-articulatory inversion.

Several studies have applied MRI for articulatory-related speech technologies: e.g.\ phoneme classification from articulation~\cite{Saha2018,VanLeeuwen2019} and acoustic-to-articulatory inversion~\cite{Kaburagi2015,Li2015}.
Saha and his colleagues experimented with identifying different vowel-consonant-vowel (VCV) sequences from vocal tract rtMRI~\cite{Saha2018}. Long-term Recurrent Convolutional Networks (including a pretrained ResNet50) models were used, which make the network spatiotemporally deep enough to capture the sequential nature of the articulatory data, for the classification task.
Van Leeuwen et al.\ trained a convolutional neural network (CNN) for the classification of 27 different sustained phonemes from MRI, and reveal that the network has learned to focus on those parts of the images that represent the crucial articulatory positions needed to distinguish the different phonemes~\cite{VanLeeuwen2019}.

Kaburagi presented a method to estimate the cross-sectional area and length of the vocal tract from a speech spectrum~\cite{Kaburagi2015}. For data, a set of vocal-tract area functions were used for 10 English vowels obtained from static MRI measurements (the subject was a male native speaker of English). However, the estimation result is the parameterized vocal tract, not the MR images itself.
As the first and single inversion study having MR images as the target, Li et al.\ present a system for AAI using midsagittal rtMRI, where restricted Boltzmann machine, GMM and linear regression are applied for the mapping~\cite{Li2015}. In this task, the inputs of the machine learning models are acoustic feature vectors (24-order line spectral pairs, with a context window of 10 acoustic frames), whereas the targets are the gray value vectors of the 68$\times$68~pixel MR images. According to the results, deep architectures are able to obtain better inversion accuracy than the GMM-based method, in terms of RMSE~\cite{Li2015}. However, only a single speaker (f1) was used from the USC-TIMIT database~\cite{Narayanan2014}.


Based on this overview, rtMRI data has sporadically been used previously for direct acoustic-to-articulatory inversion. In the current paper, we train various DNNs (fully connected, convolutional, and recurrent neural networks) for acoustic-to-articulatory inverse mapping, using real-time magnetic resonance images of the vocal tract, applying the data in a speaker-specific way.

\section{Methods}

\subsection{Data}

We used two male (m1 and m2) and two female (f1 and f2) speakers from the freely available USC-TIMIT MRI database \cite{Narayanan2014}. This contains large-scale data of synchronized audio and rtMRI for speech research, from American English subjects. The vocal tracts were imaged in the mid-sagittal plane while lying supine and reading 460 MOCHA-TIMIT sentences. The MRI data were acquired using a Signa Excite HD 1.5T scanner with an image resolution in the mid-sagittal plane of 68$\times$68~pixels (2.9$\times$2.9mm).
The image data were reconstructed as 23.18~frames/second. The audio was simultaneously recorded at a sampling frequency of 20~kHz inside the MRI scanner while subjects were imaged. Noise cancellation was also performed on the acoustic data.

\subsection{Vocoder}

To create the acoustic inputs, we encoded the audio recordings using an MGLSA vocoder~\cite{Imai1983} at a frame shift of 1 / (23.18 fps) = 863 samples, which resulted in F0 and 24-order spectral (Mel-Generalized Cepstrum, Line Spectral Pair, MGC-LSP) features. The spectral features served as the training inputs of the DNN.

\subsection{Deep neural network architectures}

In our earlier studies on ultrasound-based inversion~\cite{Porras2019}, we were using fully-connected feed-forward neural networks (FC-DNN). For articulatory mapping, we showed that CNN and CNN-LSTM were also beneficial~\cite{Moliner2019}. Here, we test similar DNN types.
In all cases and for all speakers, we trained speaker-specific models, and we split the data into 430 sentences for training, 20 sentences for validation, and 10 sentences for testing. We used Adam optimizer, trained the networks for 100 epochs, and applied early stopping with a patience of 5 on the validation loss. The target MRI pixels were scaled to [0-1], while the input spectral features were normalized to zero mean, unit variance. The data is passed to the networks in batches of 128 frames. The cost function applied for the MRI pixel regression task was the mean-squared error (MSE). In all hidden layers, ReLU activation was used.

\subsubsection{FC-DNN (baseline)}

In the simplest case, we trained fully connected (FC) DNNs with 5 hidden layers, each hidden layer consisting of 1000 neurons. The input layer consisted of 25 neurons (taking the MGC-LSP dimensions). The output layer was a linear one (4\,624 neurons), with one neuron for each MRI pixel.

\subsubsection{CNN}
\label{sec:CNN}

Next, we tested convolutional neural networks, as typically, they are more suitable for the processing of images than simple FC-DNNs. The CNN input was 25-dimensional MGC-LSP.
First, two dense layers were used with 500 and 17$\times$17$\times$8 = 2312 neurons, respectively. There was one convolutional layer (kernel size: 3$\times$3, number of filters: 8), preceded and followed by max-pooling. The last layer was a convolutional one with linear activation and 68$\times$68 pixel MR image as the target.

\subsubsection{LSTM}

\begin{figure}
\centering
\includegraphics[trim=0.0cm 21.2cm 11.0cm 0.2cm, clip=true, width=0.8\columnwidth]{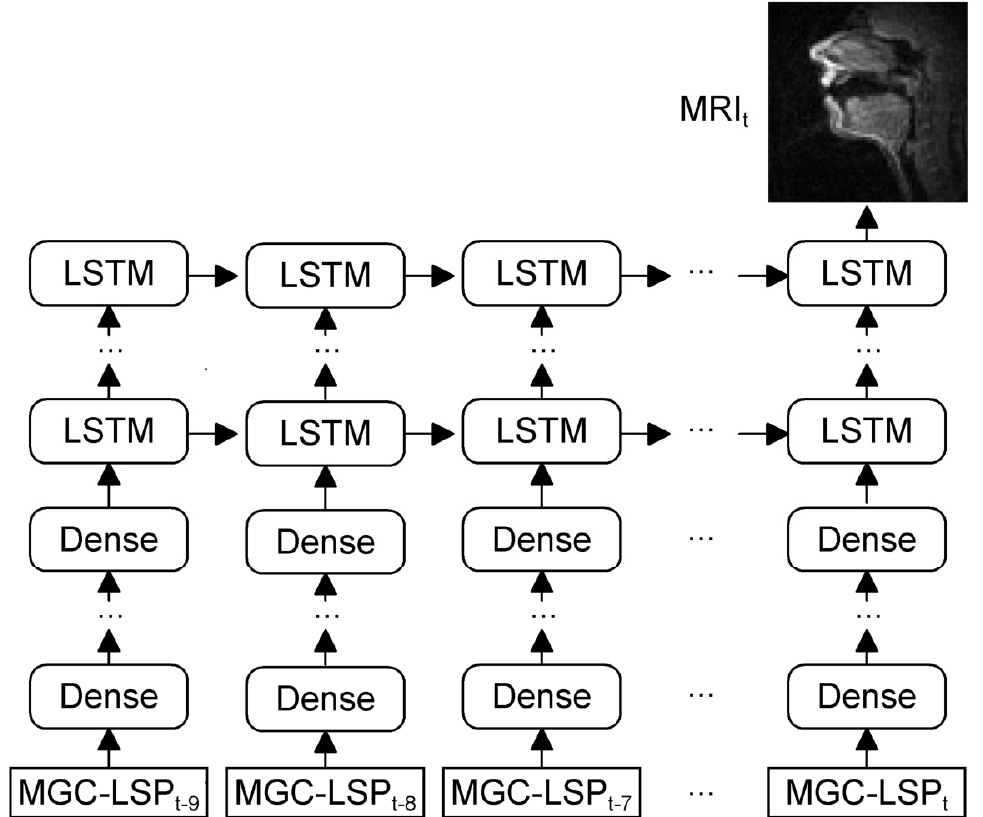}
\vspace{-1mm}
\caption{Block diagram of the LSTM network.}
\label{fig:lstm}
\vspace{-4mm}
\end{figure}

Also, we hypothesized that using multiple consecutive spectral features as input can increase the accuracy of the regression. 
The most ambitious network in this work is a recurrent one consisting of a combination of FC layers and Long Short-Term Memory units (LSTMs). The motivation for designing this network comes from the fact that in \cite{Moliner2019}, we achieved better results when using consecutive ultrasound frames.
Figure \ref{fig:lstm} shows the architecture of the LSTM network. It consists of two distinct parts: a fully connected beginning (three fully connected layers with 575 neurons each), and a recurrent end (two LSTM layers with 575 neurons each). We use a sequence size of 10 (accounting for roughly 430~ms of the input spectral data) in order to incorporate time information. 
To keep the FC-DNN (baseline) and the LSTM comparable with respect to parameter count, both models have approximately 8.6 million tunable parameters -- this was the reason for having 575 neurons in the layers of the recurrent neural network.


\begin{figure*}
\centering
\includegraphics[trim=0.0cm 0.0cm 0.0cm 0.0cm, clip=true, width=\textwidth]{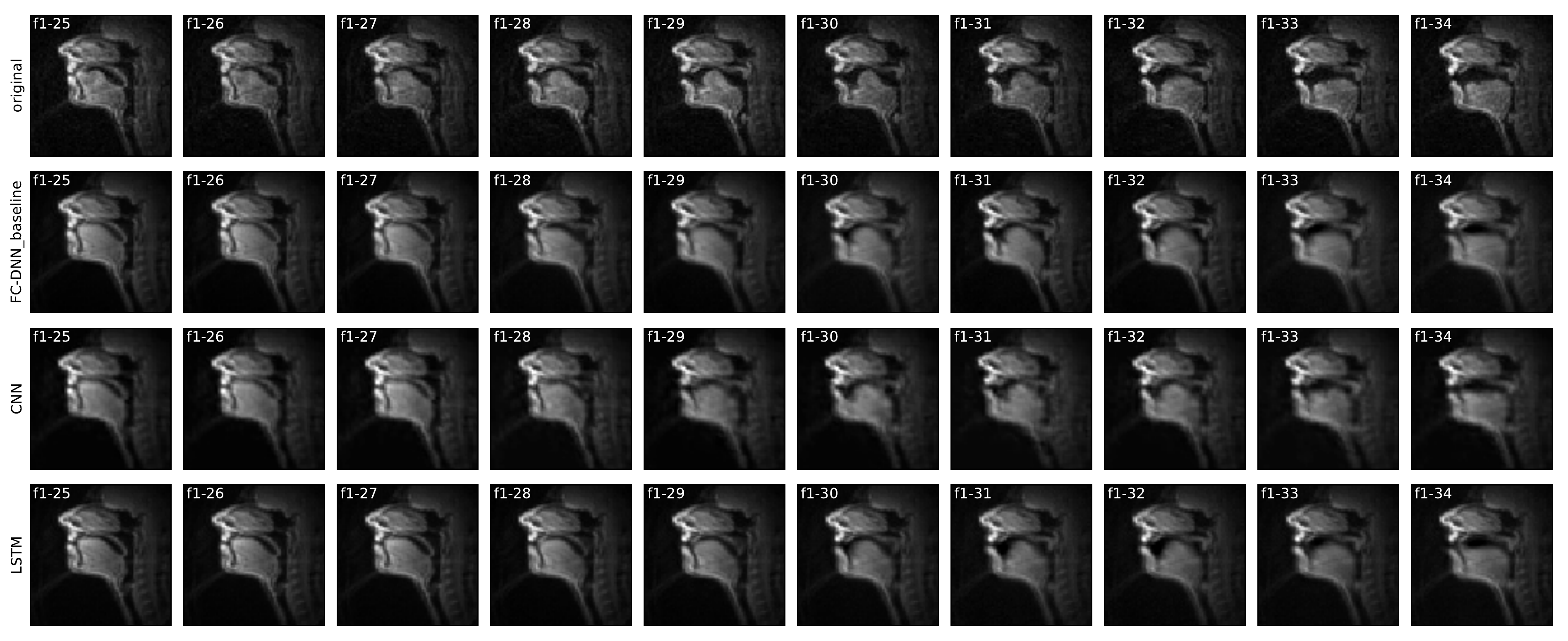}
\caption{Original and predicted MRI image sequence by the three DNNs, from speaker f1. The sentence starts with `\textit{Her...}' (f1\_146)}
\label{fig:proposed_sample}
\end{figure*}

\section{Results}

After training the neural networks, the prediction accuracy was evaluated on the test set (10 sentences for each speaker). We generated MR image sequences using the trained DNN models by having the MGC-LSP features of the test sentences as input.

\subsection{Demonstration sample}

A sample test sentence from speaker 'f1' (not appearing in the training data) was chosen for demonstrating how the three systems deal with the prediction of the MRI pixels. Fig.~\ref{fig:proposed_sample} shows the same image sequence from the original recording and the predictions of the DNNs. In general, the DNN-predicted MR images are close to the original, but the FC-DNN and CNN cannot follow the fast changes of the reference, while the LSTM seems to be smoother. More video samples can be found at \url{http://smartlab.tmit.bme.hu/interspeech2020_speech2mri}.

\subsection{Objective evaluation}
\label{sec:objective}

On the validation set and on the synthesized sentences (being the test set), we first measured the Mean Square Error (MSE) between the original and predicted MRI pixels. The calculations were done on the [0-1] normalized features. The normalized MSE values calculated on the validation and test are shown in Table~\ref{tab:objective_NMSE}, separately for each speaker. Overall, the tendencies are the same for all speakers: the weakest network seems to be the basic fully-connected one (test NMSE: 0.0043), followed by the CNN (test NMSE: 0.0049), and finally, the LSTM having the smallest error (test NMSE: 0.0036). The difference between the FC-DNN and CNN is small, whereas the LSTM is significantly better than these two systems. Interestingly, the results show some speaker dependency: the test NMSE is smallest for f1 (0.0023), and highest for f2 (0.0046), thus being roughly twice as bad. Another observation is that for speaker m1, the LSTM network is not better than the FC-DNN and CNN.

\begin{table}
\caption{NMSE scores on the validation and test set.} \label{tab:objective_NMSE}
\centering
\begin{tabular}{l||c|c|c}
        & \multicolumn{3}{c}{{Normalized MSE (validation)}} \\
\cline{2-4}
speaker & ~~FC-DNN~~ & ~~~CNN~~~ & ~~LSTM~~ \\
\hline\hline
f1 & 0.0024 & 0.0026 & 0.0021   \\
f2 & 0.0041 & 0.0042 & 0.0038   \\
m1 & 0.0028 & 0.0030 & 0.0026   \\
m2 & 0.0031 & 0.0034 & 0.0028   \\
\hline
average & 0.0031 & 0.0033 & 0.0029   \\

\hline\hline

speaker        & \multicolumn{3}{c}{{Normalized MSE (test)}} \\
\cline{2-4}
\hline\hline
f1 & 0.0028 & 0.0032 & 0.0023   \\
f2 & 0.0060 & 0.0065 & 0.0046   \\
m1 & 0.0035 & 0.0039 & 0.0035   \\
m2 & 0.0050 & 0.0058 & 0.0038   \\
\hline
average & 0.0043 & 0.0049 & 0.0036   \\

\end{tabular}

\end{table}

To measure the naturalness of the reconstructed MR images, two other metrics were chosen, which have a higher correlation with subjective quality. Both of these are calculated over each frame, where \(y\) is the original image  and \(\hat{y}\) is the estimated image from the DNN architecture.
Structural Similarity Index (SSIM)~\cite{Wang2004} (which we used earlier to compare ultrasound tongue images~\cite{Xu2016a}) measures three kinds of visual impact of changes in luminance \(l\), contrast \(c\) and structure \(s\) between two images:
\[
SSIM(y,\hat{y}) = [l(y,\hat{y})]^\alpha[c(y,\hat{y})]^\beta[s(y,\hat{y})]^\gamma
\]

In our experiment, the SSIM index is calculated by 11$\times$11 circular-symmetric Gaussian weighting function, with a standard deviation of 1.5~pixels.

Complex Wavelet Structural Similarity (CW-SSIM)~\cite{Sampat2009} is an extension of the SSIM method to the complex wavelet domain, which is a novel image similarity measurement robust to small distortions:
\[
CW-SSIM(y,\hat{y}) = \frac{2\arrowvert\sum_{l=1}^{L}w_{y,l}w_{\hat{y},l}^{*}\arrowvert + K}{\sum_{l=1}^{L}|w_{y,l}|^2 + \sum_{l=1}^{L}|w_{\hat{y},l}|^2 + K}
\]

where \(w\) represents the complex wavelet coefficients of the two images. The \(^*\) indicates the complex conjugate of \(w\), and \(K\) is a small positive stabilizing constant~\cite{Xu2016a}. In case of both SSIM and CW-SSIM, the resulting range is between [0-1], and the higher value means more similar images (whereas zero is for the most diverse images).

\begin{figure}
\centering
\includegraphics[trim=0.26cm 0.3cm 0.32cm 0.36cm, clip=true, width=\columnwidth]{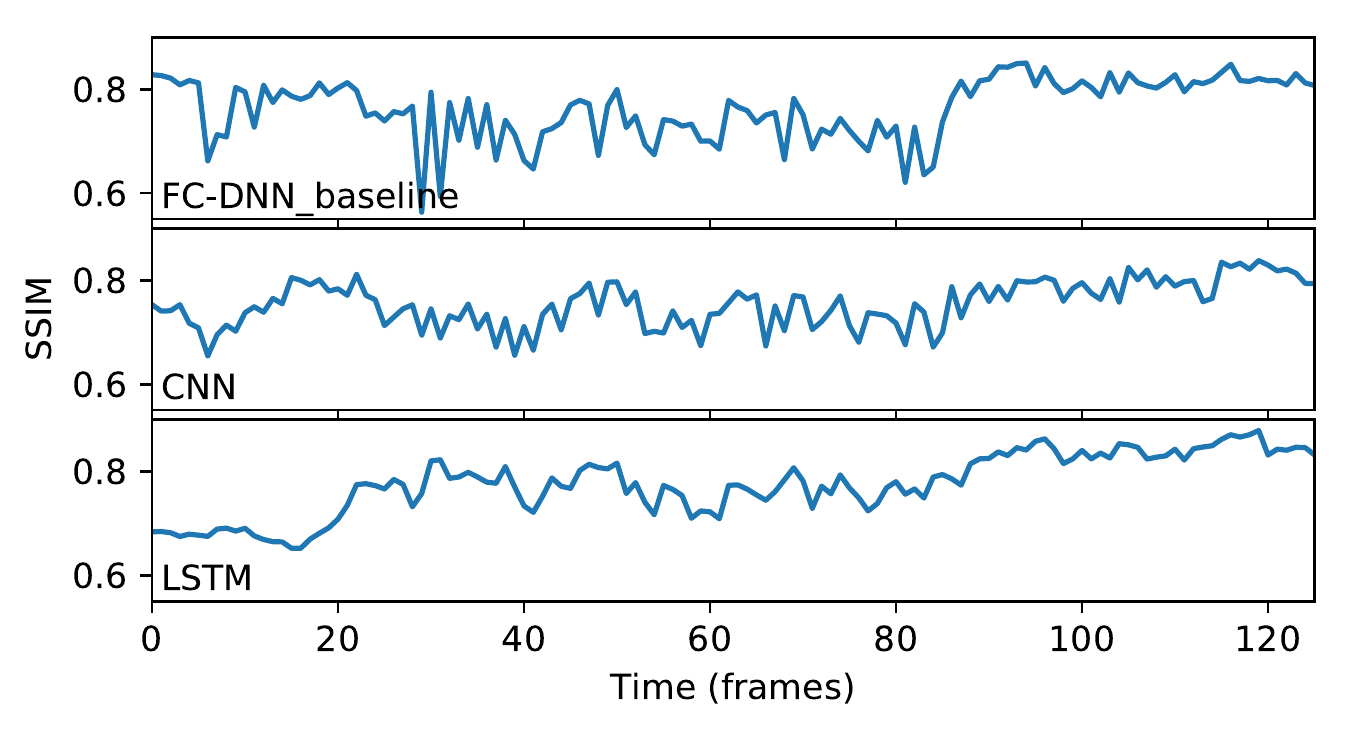}
\includegraphics[trim=0.26cm 0.3cm 0.32cm 0.25cm, clip=true, width=\columnwidth]{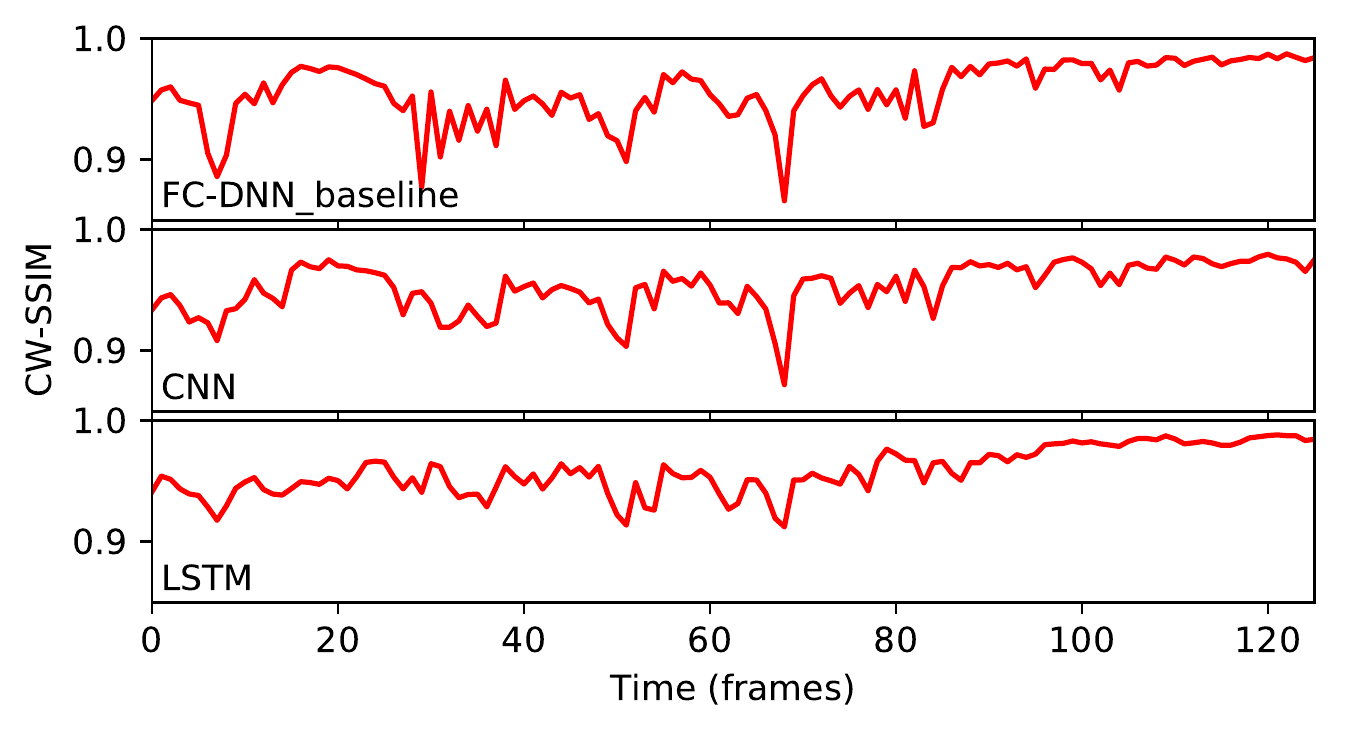}
\caption{\textit{SSIM and CW-SSIM over time, from speaker f1. The utterance, between frames 28--92 is `\textit{Her auburn hair reminded him of autumn leaves.}' (f1\_146)}}
\label{fig:objective_SSIM}
\vspace{-5mm}
\end{figure}

Table~2 shows the SSIM and CW-SSIM scores measured on the synthesized sentences, separately for each speaker and DNN type. The scores are not as speaker dependent as it was the case with NMSE. The tendencies across the neural network type show that the CNN achieved the lowest scores, followed by the FC-DNN, and finally, the LSTM having the most similar predicted images to the original MR frames; but all of the differences are small and not significant.
Fig.~\ref{fig:objective_SSIM} presents SSIM and CW-SSIM over time, on the f1\_146 sample sentence. The utterance starts around frame 28 and ends around frame 92. Frames 1--27 and 93--125 are silence. The figure shows that the FC-DNN and CNN can create videos that have quick variations across consecutive frames (e.g.\ between frames 28--40, where the sentence starts), whereas the result of LSTM is smoother. In general, the MR frames for the silence period are most similar to the original MR images, while the LSTM can achieve high scores even in case of the parts where the articulators are moving, which is clearly beneficial.

According to these objective experiments, all measures have shown the advantage of using recurrent networks (namely, LSTM), instead of the networks which are taking single images as input (the FC-DNN and CNN types).

\begin{table}

\setlength\tabcolsep{5pt}

\caption{SSIM and CW-SSIM scores on the test set.} \label{tab:objective_SSIM}
\vspace{-2mm}
\centering
\begin{tabular}{l||c|c|c}
        & \multicolumn{3}{c}{{SSIM / CW-SSIM}} \\
\cline{2-4}
speaker & ~~FC-DNN~~ & ~~~CNN~~~ & ~~LSTM~~ \\
\hline\hline
f1 & 0.77 / 0.94 & 0.75 / 0.94 & 0.80 / 0.95   \\
f2 & 0.70 / 0.92 & 0.68 / 0.92 & 0.73 / 0.94   \\
m1 & 0.80 / 0.92 & 0.77 / 0.91 & 0.81 / 0.92   \\
m2 & 0.75 / 0.92 & 0.73 / 0.92 & 0.77 / 0.94   \\
\hline
average & 0.76 / 0.93 & 0.73 / 0.92 & 0.78 / 0.94   \\

\end{tabular}
\vspace{-3mm}

\end{table}

\section{Discussion}

In general, recurrent neural networks are more suitable to process sequential data than convolutional or simple fully-connected networks, and they can generate more smooth consecutive images. As Saha et al.\ compare for MRI-based phone recognition~\cite{Saha2018}, it is important to incorporate spatial and temporal feature extraction steps to capture complementary information from the individual consecutive still frames as well as between the frames. In the current study, we showed that LSTMs are more suitable to process spectral features and MR images than FC-DNNs and (2D) CNNs.

Li and his colleagues presented the first study in SD-AAI with rtMRI of the vocal tract as the target~\cite{Li2015}. With stacked RBMs and Gaussian-Bernoulli RBMs they achieved an average RMSE of 17.74 (measured on the unnormalized pixels). However, when checking the generated images, we can see that the shape of the vocal tract is over-smoothed. They indicate that in the resulting MR images, there is a relatively higher error in the outline of nose which is believed to be caused by the instability of head position in the USC-TIMIT database.

We can compare the results of this study to acoustic-to-articulatory inversion experiments that were using other imaging techniques. Ultrasound can only capture the movement of the tongue, but with higher frame rates (around 100~fps).
In our earlier experiment for speech-to-ultrasound conversion, the typical values of SSIM were around 0.7 and CW-SSIM around 0.8, with a 2-layer FC-DNN for a single female speaker~\cite{Porras2019}, whereas for rtMRI with the LSTM here we achieved SSIM between 0.73--0.81 and CW-SSIM between 0.92--0.95, depending on the speaker.
Wei et al.\ used GMMs and DNNs for ultrasound-based AAI and achieved (unnormalized) MSE around 32--35, which is not directly comparable to our results~\cite{Wei2016}. However, their study was quite limited as they only focused on Chinese vowels, i.e.\ they did not test longer speech.
  
Although the resolution of the target MR images was only 68$\times$68~pixels, this accounts for a larger 'relative' spatial resolution compared to ultrasound, as MRI can show the whole vocal tract. Our experiments have indicated that this is an advantage of rtMRI, and the high relative spatial resolution is more important than the relatively low (around 20--25~fps) time resolution.

\section{Conclusions}

In this work, we used midsagittal rtMRI images of the vocal tract for speaker dependent acoustic-to-articulatory inversion. We applied FC-DNNs, CNNs, and recurrent neural networks and have shown that LSTMs are the most suitable for this task.

The target realtime MR images have a relatively low spatial and temporal resolution (but high 'relative' spatial resolution) and are infested with noises and reconstruction artifacts~\cite{Saha2018}. In our work, we were using raw MR images and did not apply any preprocessing. However, noise and artifact reduction on the target images, or other spectral feature extraction methods~\cite{Illa2019b} might enhance the accuracy of the mapping. Stabilizing the head position on the MR images can also be beneficial~\cite{Li2015}.

As pointed out in Section 1, the results in AAI might be useful for speech recognition~\cite{King2007}, synthesis~\cite{Ling2009}, talking heads~\cite{Wang2010}, and for pronunciation training and language tutoring~\cite{Katz2014}.

The keras implementations are accessible at \url{https://github.com/BME-SmartLab/speech2mri/}.

\section{Acknowledgements}

The author was funded by the National Research, Development and Innovation Office of Hungary (FK 124584 and PD 127915 grants). The Titan X GPU for the deep learning experiments was donated by the NVIDIA Corporation. We would like to thank USC for providing the USC-TIMIT articulatory database.

\clearpage

\bibliographystyle{IEEEtran}

\bibliography{ref_collection_csapot_nourl}

\end{document}